\documentclass[10pt]{iopart}
\pdfoutput=1
\usepackage{iopams}  

\expandafter\let\csname equation*\endcsname\relax
\expandafter\let\csname endequation*\endcsname\relax

\usepackage{amssymb}
\usepackage{amsmath}

\usepackage{soul}
\usepackage{cite}
\usepackage{enumerate}

\usepackage{empheq}

\usepackage{fancybox}

\usepackage{etoolbox}
\usepackage{hyperref}

\usepackage{color}

\makeatletter
\def\@mkboth#1#2{}
\newlength\appendixwidth
\preto\appendix{\addtocontents{toc}{\protect\patchl@section}}
\newcommand{\patchl@section}{%
  \settowidth{\appendixwidth}{\textbf{Appendix }}%
  \addtolength{\appendixwidth}{1.5em}%
  \patchcmd{\l@section}{1.5em}{\appendixwidth}{}{\ddt}%
}
\makeatother

\begin{document}


\title[Covariance of the running range]{Covariance of the running range of a Brownian trajectory}

\author{Brandon Annesi$^{1,2}$, Enzo Marinari$^{1,2}$ \& Gleb Oshanin$^3$}
\address{$^1$ Dipartimento di Fisica,
  Sapienza Universit{\`a} di Roma, P.le A. Moro 2, I-00185 Roma,
  Italy}
\address{$^2$ INFN, Sezione di Roma 1 and Nanotech-CNR, UOS di
  Roma, P.le A. Moro 2, I-00185 Roma, Italy}
\address{$^3$ Sorbonne Universit\'e, CNRS, Laboratoire de Physique Th\'eorique de la Mati\`ere Condens\'ee (UMR 7600),
 4 Place Jussieu, 75252 Paris Cedex 05, France}
\eads{brandonlivio@gmail.com, enzo.marinari@uniroma1.it,  oshanin@lptmc.jussieu.fr}

\begin{abstract}
The question how the extremal values of a stochastic process
achieved on different time intervals are correlated to each other 
has been discussed within the last few years on examples of the running maximum of a Brownian motion, of a Brownian Bridge and of a Slepian process. 
Here, 
we focus on the two-time correlations of 
the running \textit{range} of Brownian motion - the maximal extent of a Brownian trajectory 
on a finite time interval. We calculate exactly the covariance function of the running range and analyse its asymptotic behaviour.  
Our analysis reveals non-trivial correlations between  the value of the
 largest descent (rise) of a BM
 from the top to a bottom on some time interval, 
 and the value of this property 
 on a larger time interval. 
\end{abstract}

\vspace{2pc}

\vspace{2pc}



\maketitle


\section{Introduction} 

Brownian motion (BM)  is a paradigmatic stochastic
process encountered in diverse areas
of physics
and chemistry \cite{2,frey},  biology \cite{frey,4}, 
differential evolutionary games \cite{5,6,7},
computer science \cite{8,80}, as well as in 
mathematical finance \cite{9,10,11}, 
where it represents one of the main components
in modelling of the dynamics of asset prices. 

\begin{figure}[ht]
\centering
\includegraphics[width=0.5\textwidth]{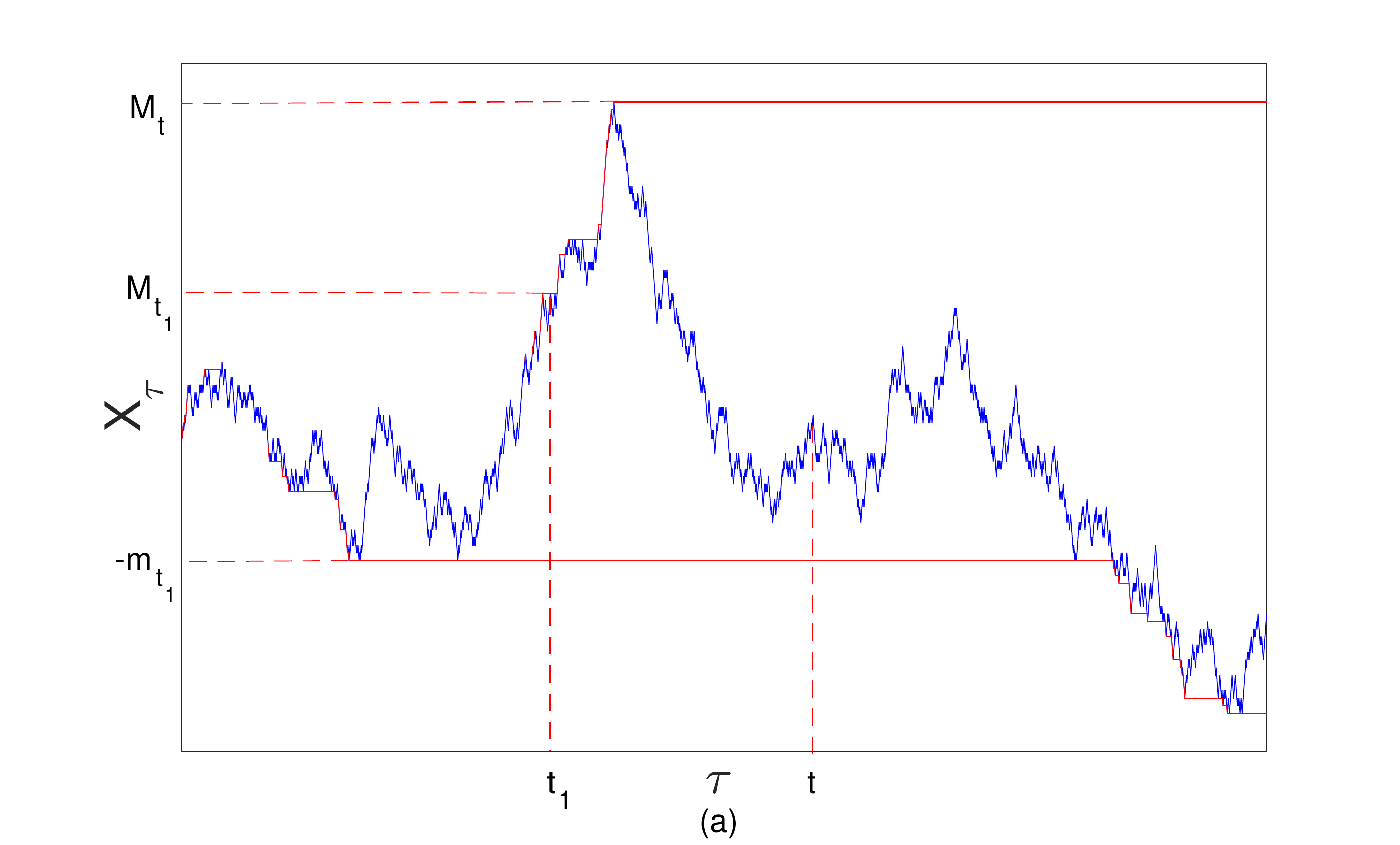}
\includegraphics[width=0.48\textwidth]{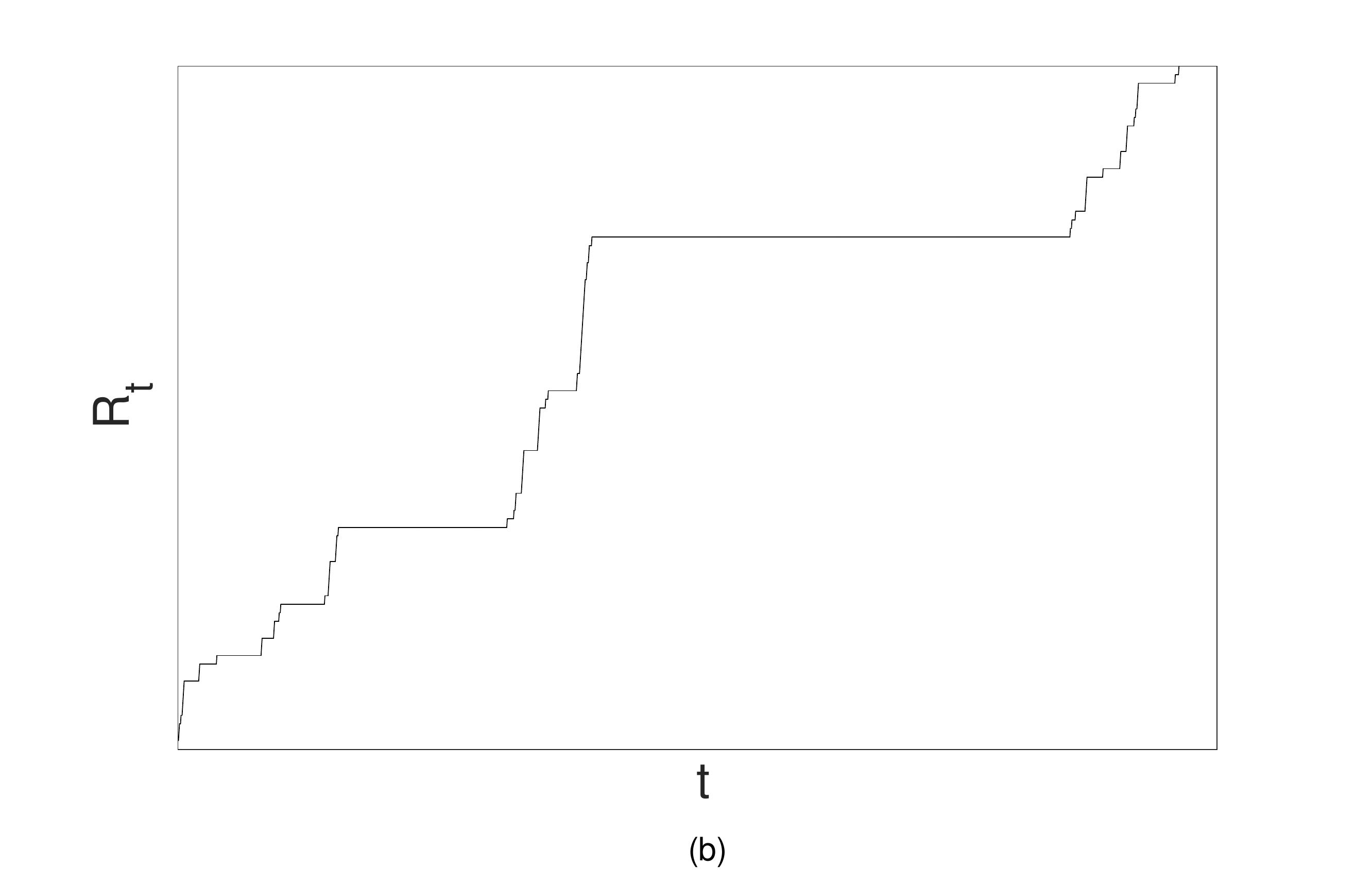}
\caption{Panel (a): Maximal and minimal displacements of a Brownian trajectory $X_{\tau}$ on time intervals $[0,t_1]$ and $[0,t]$. Panel (b): Running range of a given BM trajectory $X_{\tau}$ depicted in panel (a).}
\label{FIG1}
\end{figure}

Extremal values of a BM, e.g., its
maximal and minimal displacements, 
its range (see Fig. \ref{FIG1}), 
or
first passage times to a given level or a given point in space,  are known to 
play a very important role across many disciplines, since they sometimes prompt or 
trigger some 
particular response of the system  \cite{1,100,41,45,ol,3,24,pccp}. 
Since the early seminal works \cite{1,kol,smi,fel}, 
much effort has been 
invested into the analysis of such  extreme values. 
The distributions of these properties themselves, 
or even
more complicated 
joint multivariate distributions have been determined 
exactly \cite{1,100,kol,smi,fel,greg} or using some accurate approximate approaches (see, e.g., recent Ref. \cite{pccp}). 
Knowledge of these distributions is often very useful for a non-perturbative analysis 
of complicated functionals of BM, permitting 
to construct convergent bounds on these functionals
and hence, to obtain non-trivial exact results 
 \cite{bound0,bound1,bound2,bound20,bound03,bound3}.

In contrast, the question how the extremes of BM are correlated in time 
has received essentially less attention. 
Only within the last few years,
the two-time correlations of the running maximum of BM \cite{111}, of a Brownian Bridge (a BM constrained to return to the starting point at a fixed time moment $t$)
\cite{112} , the temporal correlations of a Slepian process (the difference of two BM positions taken at different time moments) \cite{113}
and of some records related to BM \cite{114,115}
have been determined. 
Here we focus on the two-time correlations (the covariance function) of the running range of a Brownian motion (see Fig. \ref{FIG1})
  - a maximal extent (span) of a BM on a given time interval. Our analysis reveals non-trivial correlations between
the value of the
 largest descent (rise) of a BM
 from the top to a bottom on some time interval, 
 and the value of this property 
 on a different (larger) time interval. 
 
 This paper is outlined as follows: in Sec. \ref{model} we introduce basic notations and present our main results. In Sec. \ref{details},  we present the details of the derivations. Finally, in Sec. \ref{conc} we conclude with a brief recapitulation of our results.

\section{Notations and the main results}
\label{model}

Let $X_{\tau}$, $X_{\tau =0}$,  denote the position of a one-dimensional BM with diffusion coefficient $D$ 
at time moment $\tau$, $0 \leq \tau \leq t$, and let
\begin{align}
\label{1}
M_t = {\rm max}_{0 \leq \tau \leq t} X_{\tau}  \geq 0 \, \,, m_t = - {\rm min}_{0 \leq \tau \leq t} X_{\tau} \geq 0 \,, \, \textrm{and} \,\, R_t = M_t + m_t  \geq 0 \,,
\end{align}
denote the maximal positive displacement $M_t$ of $X_{\tau}$ within the interval $0 \leq \tau \leq t$, 
the minimum (i.e.,  the maximal negative displacement taken with a negative sign) $m_t$ of $X_{\tau}$ within this time interval, and the range 
$R_t$
of $X_{\tau}$ within this interval, respectively (see Fig. \ref{FIG1}).  Our goal is to calculate exactly the covariance function of the running range $R_t$, which is defined as
$\mathbb{E}\left\{R_{t_1} R_{t}\right\} $, $0 \leq t_1 \leq t $, where the symbol $\mathbb{E}\left\{\ldots\right\}$ here and henceforth denotes averaging with respect to all possible realisations of $X_{\tau}$.

In virtue of  eq. \eqref{1}, the covariance function of the running range can be written down as
\begin{align}
\mathbb{E}\left\{R_{t_1} R_{t}\right\} &= \mathbb{E}\left\{\left(M_{t_1} + m_{t_1}\right)\left(M_{t} + m_{t}\right)\right\} \nonumber\\
&= \mathbb{E}\left\{M_{t_1} M_{t}\right\} + \mathbb{E}\left\{m_{t_1} m_{t}\right\} + \mathbb{E}\left\{M_{t_1} m_{t}\right\} + \mathbb{E}\left\{m_{t_1} M_{t}\right\} \,.
\end{align}
By symmetry, we have that
\begin{align}
&\mathbb{E}\left\{M_{t_1} M_{t}\right\} = \mathbb{E}\left\{m_{t_1} m_{t}\right\} \,, \,\,\, \mathbb{E}\left\{M_{t_1} m_{t}\right\} = \mathbb{E}\left\{m_{t_1} M_{t}\right\} \,,
\end{align}
such that 
\begin{align}
\label{lm}
\mathbb{E}\left\{R_{t_1} R_{t}\right\} &= 2 \Big(\mathbb{E}\left\{M_{t_1} M_{t}\right\}  +  \mathbb{E}\left\{m_{t_1} M_{t}\right\}\Big) \,.
\end{align}
The first term in the brackets in the latter equation is known \cite{111}. In contrast, the expected value of the product of a minimum and a maximum of a BM attained at two different time intervals
has not been yet determined. Our first main result is the following exact representation
\begin{align}
\label{chief1}
& \frac{\mathbb{E}\left\{m_{t_1} M_{t}\right\}}{D t}  =  \frac{4}{ \pi} \sqrt{z (1 -z)} - 1  - z
 + \frac{2}{\pi} \frac{z^{3/2}}{\sqrt{1 - z}} g_1\left(\sqrt{\frac{z}{1-z}}\right)+ \nonumber\\
 &+ \frac{1}{\pi} \left[\left(\sqrt{\frac{z}{1-z}} + i\right) g_1\left(\sqrt{\frac{z}{1-z}} + i\right) + \left(\sqrt{\frac{z}{1-z}} - i\right) g_1\left(\sqrt{\frac{z}{1-z}}-i\right) 
 \right] \,,
\end{align}
where $i = \sqrt{-1}$, $z =t_1/t$, and the function $g_1(\phi)$ is defined by
\begin{align}
\label{gg1}
g_1(\phi) & = \frac{1}{\phi^2} \int^1_0 \frac{d\beta}{\beta^2} \left(1 - \frac{\pi \phi \beta}{\sinh(\pi \phi \beta)}\right) \,.
\end{align}
Large- and small-$\phi$ asymptotic behavior of $g_1(\phi)$ is discussed below (see eqs. \eqref{small} and \eqref{as}). 

\begin{figure}[ht]
\begin{center}
\centerline{\includegraphics[width = .6 \textwidth]{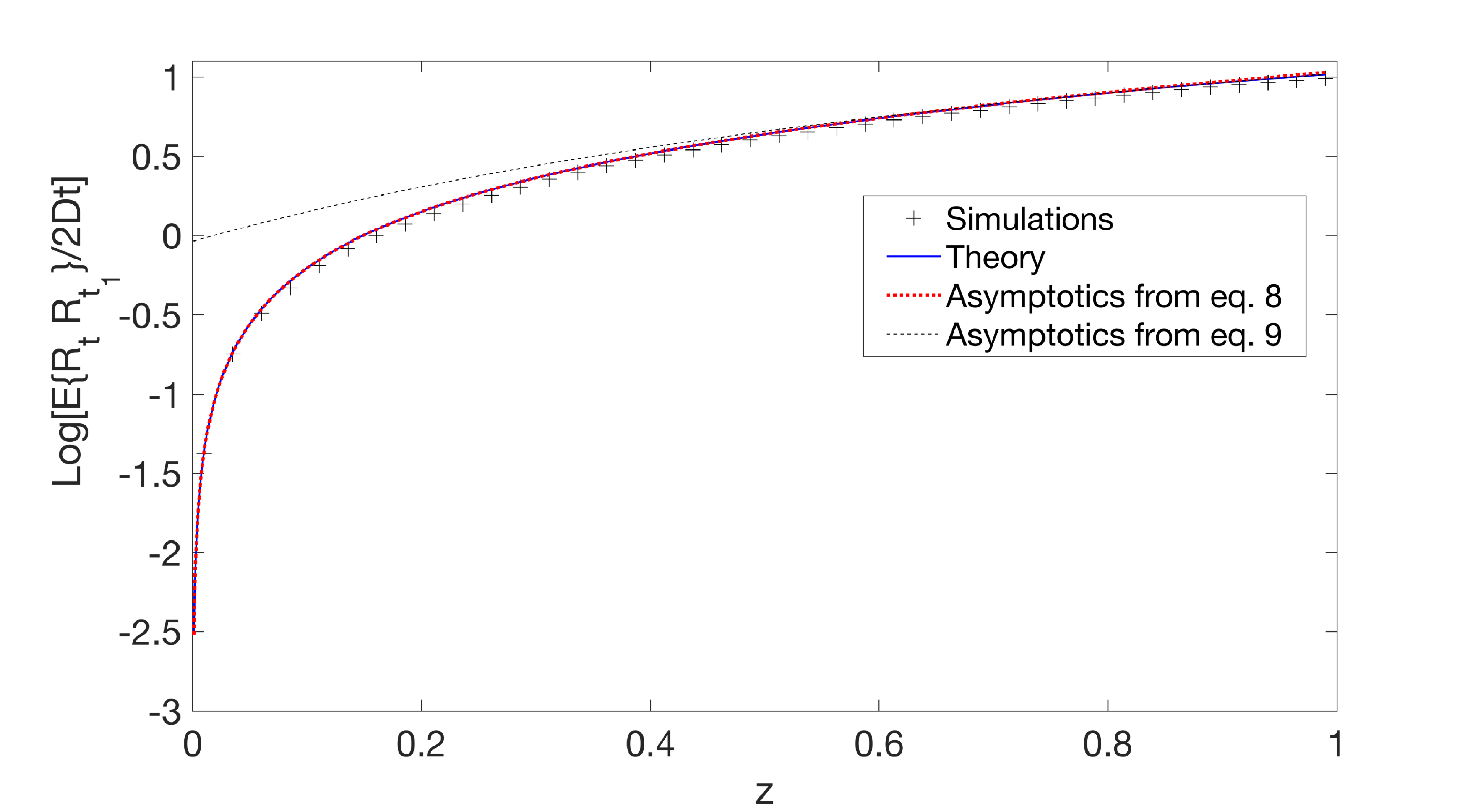}}
\caption{Exact expression for the covariance of the running range of BM in eq. \eqref{chief2} (solid blue curve), 
together with the asymptotic forms in eqs. \eqref{chief3} (red dotted curve) and \eqref{chief4} (dotted curve). The crosses are the results of numerical simulations.
\label{FIG2}
}
\end{center}
\end{figure} 

Recalling next the exact expression for  $\mathbb{E}\left\{M_{t_1} M_{t}\right\} $, derived in Ref. \cite{111}, we write down
the second main result of work, which represents the desired covariance function of the running range of BM :
\begin{align}
\label{chief2}
&\frac{\mathbb{E}\left\{R_{t_1} R_{t}\right\}}{2 D t} =   \frac{6}{ \pi} \sqrt{z (1 -z)} - 1 + \frac{2}{\pi} \arcsin\left(\sqrt{z}\right)
 + \frac{2}{\pi} \frac{z^{3/2}}{\sqrt{1 - z}} g_1\left(\sqrt{\frac{z}{1-z}}\right)+ \nonumber\\
 &+ \frac{1}{\pi} \left[\left(\sqrt{\frac{z}{1-z}} + i\right) g_1\left(\sqrt{\frac{z}{1-z}} + i\right) + \left(\sqrt{\frac{z}{1-z}} - i\right) g_1\left(\sqrt{\frac{z}{1-z}}-i\right) 
 \right] \,.
\end{align}
It is instructive to consider the asymptotic behaviour of the expression in eq. \eqref{chief2} in the limits when $t_1 \ll t$ ($z \to 0$) 
and when $t_1$ is close to $t$, such that $z \to 1$. In the former limit (see the expansions in eqs. \eqref{azer1} and \eqref{azer2} below) we obtain 
\begin{align}
\label{chief3}
&\frac{\mathbb{E}\left\{R_{t_1} R_{t}\right\}}{\mathbb{E}\left\{R_{t_1}\right\} \mathbb{E}\left\{R_{t}\right\}} = 1 +  \frac{(\pi^2 - 6)}{36} z - \frac{(270 + 7 \pi^4 - 90 \pi^2)}{10800} z^2 + O\left(z^3\right)
\end{align}
where $\mathbb{E}\left\{R_{t}\right\} = 2 \mathbb{E}\left\{M_{t}\right\} = 4 \sqrt{D t/\pi}$.  Equation \eqref{chief3} demonstrates that correlations between $R_{t_1}$ and  $R_{t}$ vanish when $t_1/t \to 0$, (which happens, namely, when $t_1$ is fixed and $t \to \infty$), but this 
limit is approached via a slow power law, which reveals long-ranged temporal correlations between the values of the running range achieved on two different time intervals. 

Within the opposite limit $z \to 1$, we find, by taking advantage of the asymptotic expansion in eq. \eqref{as}, 
\begin{align}
\label{chief4}
\frac{\mathbb{E}\left\{R_{t_1} R_{t}\right\}}{\mathbb{E}\left\{R^2_{t}\right\}}  = 1 - \frac{\left(1 - z\right)}{2} - \frac{\left(1 - z\right)^{3/2}}{3 \pi \ln(2)} + O\left(\left(1-z\right)^{5/2}\right) \,, 
\end{align} 
where $\mathbb{E}\left\{R^2_{t}\right\} = 8 \ln(2) D t$.

The asymptotic forms in eqs. \eqref{chief3} and \eqref{chief4} are presented in Fig. \ref{FIG2} together with the exact result in eq. \eqref{chief2} (solid blue curve). We observe that the small-$z$ asymptotic form in eq. \eqref{chief3} provides a very accurate estimate of the covariance function of the running range over the whole domain of variation of $z$, and only slightly overestimates the actual value of $\mathbb{E}\left\{R_{t_1} R_{t}\right\}$ in the vicinity of $z = 1$. In turn, the asymptotic form in eq. \eqref{chief3}, which describes the behaviour of $\mathbb{E}\left\{R_{t_1} R_{t}\right\}$ in the vicinity of $z=1$, is almost indistinguishable from  $\mathbb{E}\left\{R_{t_1} R_{t}\right\}$ for $z$ as low as $0.5$.

Details of the derivation of the results in eqs. \eqref{chief1}, \eqref{chief2}, \eqref{chief3} and \eqref{chief3} are presented in the next section.

\section{Details of calculations.}
\label{details}

We start with the expression for the probability  
$P\left(m | M\right)$ 
that a BM, commencing at the origin at $t=0$, has reached  on the time interval $[0,t_1]$ a minimum whose \textit{absolute} value $= m > 0$ and on the entire interval $[0,t]$ (with $t_1 \leq t$) a maximum $= M >0$ :
\begin{align}
\label{eq}
P\left(m | M\right) &= \int_{-m}^M dX \, \Pi_{t_1}\left(M |m | X \right) \, S_{t-t_1}\left(M - X\right) + \nonumber\\
&+ \int^M_{-m}  dX \, S^*_{t_1}\left(M| m | X \right) \, \Pi_{t - t_1}\left(M - X\right) \,,
\end{align}
where  \\
\begin{itemize}
\item $S^*_{t}\left(M| m | X \right)$ is the conditional "survival" probability, i.e., the probability that a BM, starting at the origin at $t=0$, attained within the time interval $[0,t]$ a minimum whose absolute value $= m$, stayed below a level $M > 0$ within this interval, and at time moment $t$
appeared at position $X$ (such that $-m < X < M$),\\
\item $\Pi_{t}\left(M| m | X \right)$ is the trivariate pdf (see Ref. \cite{1}) of the event that a BM, starting at the origin at $t=0$, attained a maximal value $= M >0$ and a minimal value $- m$ within the time interval $[0,t]$ and ended up at position $X$ ($-m < X < M$) at time moment $t$\\
\item $S_{t}\left(M\right)$ is the standard "survival" probability (see, e.g., Ref. \cite{3}), i.e., the probability that a BM starting at the origin at time moment $t = 0$, did not reach (stayed below) a level $M >0$ within the time interval $[0,t]$ \\
\item $ \Pi_{t}\left(M\right)$ is the pdf (see, e.g., Ref. \cite{1}) of the event that a BM, starting at the origin at $t = 0$, attained a maximal value $= M > 0$
on a time interval $[0,t]$.
\end{itemize}

Let now $S_t(M|m|X)$ denote the standard "survival" probability, i.e., the probability that a BM, starting at the origin, does not leave a stripe $[-m,M]$ within time interval $(0,t)$ and its endpoint at time $t$ is $X$, ($-m < X < M$). Clearly, we have
\begin{align}
S^*_{t_1}(M|m|X) = \frac{\partial }{\partial m} S_{t_1}(M|m|X)
\end{align}
and
\begin{align}
\Pi_{t_1}(M|m|X) = \frac{\partial^2 }{\partial m \partial M} S_{t_1}(M|m|X)
\end{align}
Noticing next that
\begin{align}
\Pi_{t - t_1}\left(M - X\right) = \frac{\partial}{\partial M} S_{t-t_1}\left(M - X\right)
\end{align}
we may rewrite eq. \ref{eq} formally as
\begin{align}
\label{eq2}
P\left(m | M\right) &= \int_{-m}^M dX \, \frac{\partial}{\partial M} \Bigg[S^*_{t_1}(M|m|X) S_{t-t_1}\left(M - X\right)\Bigg] \nonumber\\
&= \int_{-m}^M dX \, \frac{\partial^2}{\partial M \partial m} \Bigg[S_{t_1}(M|m|X) S_{t-t_1}\left(M - X\right)\Bigg] 
\end{align}
Further on, recalling that
\begin{align}
S_{t-t_1}\left(M - X\right) \equiv {\rm erf}\left(\frac{M-X}{\sqrt{4 D (t - t_1)}}\right) = \frac{2}{\pi} \int^{\infty}_0 \frac{dz}{z} e^{-4 D (t - t_1) z^2} \sin\Big(2 z \left(M - X\right)\Big)
\end{align}
we cast eq. \ref{eq2} into the form
\begin{align}
P\left(m | M\right) = \frac{2}{\pi} \int^{\infty}_0 \frac{dz}{z} e^{-4 D (t - t_1) z^2} \int_{-m}^M dX \, \frac{\partial^2}{\partial M \partial m} \Bigg[S_{t_1}(M|m|X) \sin\left(2 z \left(M - X\right)\right)\Bigg] \,.
\end{align}
Lastly, to simplify somehow our analysis, we will focus in what follows not on $P(m|M)$ itself, but rather on the distribution of an auxiliary range ${\cal R} = M + m$, where $M$ denotes the maximal displacement achieved on the interval $[0,t]$, while $m$ is the absolute value of the minimal displacement achieved on the subinterval $[0,t_1]$. This distribution is formally defined as  
\begin{align}
\label{dist}
P\left({\cal R}\right) = \int^{\infty}_0 dM \int^{\infty}_0 dm \, \delta\left(M+m - {\cal R}\right) \, P(m|M) \,.
\end{align} 
The second moment of this distribution 
will provide us the desired correlation function $\mathbb{E}\left\{m_{t_1} M_t\right\}$ (see eq. \eqref{lm}) through the relation
\begin{align}
\mathbb{E}\left\{m_{t_1} M_t\right\} = \frac{1}{2} \left(\mathbb{E}\left\{{\cal R}^2_t\right\} - \mathbb{E}\left\{M^2_t\right\} - \mathbb{E}\left\{m_{t_1}^2\right\}
\right) \,.
\end{align}

The survival probability $S_{t_1}(M|m|X)$ is well-known (see, e.g., Ref. \cite{1}) and is given explicitly by
\begin{align}
\label{eq3}
S_{t_1}(M|m|X) &= \frac{1}{\sqrt{4 \pi D t_1}} \sum_{k=-\infty}^{\infty} \Bigg[\exp\left(- \frac{\left(X - 2 k (M+m)\right)^2}{4 D t_1}\right) \nonumber\\
&- \exp\left(- \frac{\left(X -  2 M - 2 k (M+m)\right)^2}{4 D t_1}\right)\Bigg] \,.
\end{align}
For further convenience, it is however more expedient to use not the expression in eq. \eqref{eq3}, but rather its Laplace-transform over the "variable" $4 D t_1$:
\begin{align}
S_{\lambda}(M|m|X) &= \int^{\infty}_0 e^{- \lambda (4 D t_1)} S_{t_1}(M|m|X) \,.
\end{align}
Performing the Laplace transform and the summation over $k$, we get
\begin{align}
\label{eq4}
S_{\lambda}(M|m|X) &=  \frac{2}{\sqrt{\lambda}} \frac{\sinh\left(2 \sqrt{\lambda} \, m\right) \sinh\left(2 \sqrt{\lambda} \left(M - X\right)\right)}{\sinh\left(2 \sqrt{\lambda} \left(M + m\right)\right)} \,, \,\,\, X \geq 0 \,, \nonumber\\
S_{\lambda}(M|m|X) &=  \frac{2}{\sqrt{\lambda}} \frac{\sinh\left(2 \sqrt{\lambda} \, M\right) \sinh\left(2 \sqrt{\lambda} \left(m + X\right)\right)}{\sinh\left(2 \sqrt{\lambda} \left(M + m\right)\right)} \,, \,\,\, X \leq 0 \,.
\end{align}
 Inserting the expressions in eq. \ref{eq4} into eq. \ref{dist} and performing all the integrations, we get
 \begin{align}
 \label{zuz}
 P\left({\cal R}\right) = \frac{4}{\pi} \int^{\infty}_0 dz \, e^{-4 D (t - t_1) z^2}  {\cal L}^{-1}_{\lambda,4 D t_1}\Bigg[Q_1 + Q_2 + Q_3\Bigg] \,,
 \end{align}
where
\begin{align}
Q_1 &= \frac{\left(1 - e^{-2 \sqrt{\lambda} \, {\cal R}}\right) \left(1 - e^{-4 \sqrt{\lambda} \, {\cal R}}\right)}{\left(1 + e^{-2 \sqrt{\lambda} \, {\cal R}}\right)^3} \, \frac{\cos\left(2 z {\cal R}\right)}{\lambda + z^2} \,, \nonumber\\
Q_2 &= \frac{2 e^{- 2 \sqrt{\lambda} \, {\cal R}} \left(1 - e^{-2 \sqrt{\lambda} \, {\cal R}}\right)}{\left(1 + e^{-2 \sqrt{\lambda} \, {\cal R}}\right)^3} \, \frac{\sin\left(2 z {\cal R}\right)}{\sqrt{\lambda} \, z} \,, \nonumber\\
Q_3 &=  \frac{\left(1 - e^{-2 \sqrt{\lambda} \, {\cal R}}\right) \left(1 + e^{-4 \sqrt{\lambda} \, {\cal R}}\right)}{\left(1 + e^{-2 \sqrt{\lambda} \, {\cal R}}\right)^3} \, \frac{z \, \sin\left(2 z {\cal R}\right)}{\sqrt{\lambda} \, \left(\lambda + z^2\right)} \,,
\end{align} 
 and ${\cal L}^{-1}_{\lambda,4 Dt_1}[\ldots]$ denotes the operator of the inverse Laplace 
 transform with respect to $\lambda$. We note that one can rather straightforwardly obtain from eq. \eqref{zuz} the distribution of the auxiliary range in the limit $t_1 \to 0$ with $t$ kept fixed (or equivalently, $t \to \infty$ with $t_1$ kept fixed) in which ${\cal R} \to M_t$, the latter being the maximum attained 
by a BM trajectory within the time interval $[0,t]$. In this limit, corresponding to $\lambda \to \infty$, in the leading in $\lambda$ order
\begin{align}
Q_1 \sim \frac{\cos\left(2 z {\cal R}\right)}{\lambda + z^2} \,,
\end{align}
while $Q_2$ and $Q_3$ vanish. Consequently, here we have that the distribution in eq. \eqref{zuz} becomes
 \begin{align}
 \label{zuz1}
P\left({\cal R}\right) \sim \frac{4}{\pi} \int^{\infty}_0 dz \, e^{-4 D t z^2}  {\cal L}^{-1}_{\lambda,  4 D t_1} \Bigg[\frac{\cos\left(2 z {\cal R}\right)}{\lambda + z^2} \Bigg] \to \frac{1}{\sqrt{\pi D t}} \exp\left( - \frac{{\cal R}^2}{4 D t}\right) \,,
 \end{align}
which is the celebrated result due to L\'evy for the distribution of the maximum of a BM trajectory within the time interval $[0,t]$. The form of the distribution in the opposite limit when $t_1 \to t$ (and hence, the auxiliary range tends to the full range, i.e., ${\cal R} \to R_t$), will be verified below.

Finally, we seek the moments of the auxiliary range:
\begin{align}
\mathbb{E}\left\{{\cal R}^n\right\} = \int^{\infty}_0 d{\cal R} \, {\cal R}^n \, P\left({\cal R}\right) \,.
\end{align}
Changing the integration variables $z \to \sqrt{\lambda} \, \xi$ and ${\cal R} \to \rho/\sqrt{\lambda}$, we have
\begin{align}
\label{zu}
\mathbb{E}\left\{{\cal R}^n\right\} &= \frac{4}{\pi} \int^{\infty}_0 \frac{\left(1 - e^{-2 \rho}\right)}{\left(1 + e^{-2 \rho}\right)^3} \, \rho^n \, d\rho \int^{\infty}_0 d\xi  \Bigg[\left(1 - e^{-4 \rho}\right) \frac{\cos\left(2 \xi \rho\right)}{1+ \xi^2} + \nonumber\\
&+ 2 e^{- 2 \rho} \frac{\sin\left(2 \xi \rho\right)}{\xi} + \left(1 + e^{-4 \rho}\right) \frac{\xi \, \sin\left(2 \xi \rho\right)}{1+ \xi^2}\Bigg]  {\cal L}^{-1}_{\lambda,4 D t_1}\Bigg[\frac{e^{- 4 D (t - t_1) \lambda \xi^2}}{\lambda^{1+ n/2}}\Bigg]
\end{align}
The inverse Laplace transform in eq. \eqref{zu}  can now be explicitly performed,
\begin{align}
{\cal L}^{-1}_{\lambda,4 D t_1}\Bigg[\frac{e^{- 4 D (t - t_1) \lambda \xi^2}}{\lambda^{1+ n/2}}\Bigg] = 
\frac{\left(4 D t_1\right)^{n/2}}{\Gamma\left(n/2 + 1\right)} \left(1 - \frac{(t- t_1)}{t_1} \xi^2\right)^{n/2} \theta\left(1 - \frac{(t - t_1)}{t_1} \xi^2\right) \,,
\end{align}
where $\theta(x)$ is the Heaviside theta-function such that $\theta(x) = 1$ for $x \geq 0$ and is zero, otherwise.
 
Therefore, we have the following expression for the moments of the auxiliary range: 
\begin{align}
\label{kl}
&\mathbb{E}\left\{{\cal R}^n\right\} = \frac{2 (D t_1)^{n/2}}{\pi \, \Gamma\left(n/2+1\right)} \int^{\infty}_0 \frac{\left(1 - e^{- \rho}\right)}{\left(1 + e^{-  \rho}\right)^3} \, \rho^n \, d\rho \int^{\infty}_0 d\xi  \Bigg[\left(1 - e^{-2 \rho}\right) \frac{\cos\left( \xi \rho\right)}{1+ \xi^2} + \nonumber\\
+& 2 e^{-  \rho} \frac{\sin\left(\xi \rho\right)}{\xi} +  \left(1 + e^{-2 \rho}\right) \frac{\xi \, \sin\left( \xi \rho\right)}{1+ \xi^2}\Bigg]  \left(1 - \frac{(t-t_1)}{t_1} \xi^2\right)^{n/2} \theta\left(1 - \frac{(t - t_1)}{t_1} \xi^2\right) \,, 
\end{align}
which can be formally rewritten, upon an appropriate change of the integration variable as
\begin{align}
\label{kl7}
\mathbb{E}\left\{{\cal R}^n\right\}   &= \frac{2 \left(D t_1\right)^{n/2} \phi}{\pi \Gamma\left(n/2 +1\right)}  \int^{\infty}_0 \frac{\left(1 - e^{- \rho}\right)}{\left(1 + e^{-  \rho}\right)^3} \, \rho^n \, d\rho \int^{1}_0 dx \left(1 - x^2\right)^{n/2} \nonumber\\
\times&  \Bigg[\left(1 - e^{-2 \rho}\right) \frac{\cos\left(\phi x \rho\right)}{1+ \phi^2 x^2} +  2 e^{-  \rho} \frac{\sin\left(\phi x  \rho\right)}{\phi x} +  \left(1 + e^{-2 \rho}\right) \frac{\phi x  \, \sin\left( \phi x \rho\right)}{1+ \phi^2 x^2}\Bigg] \nonumber\\
=& \frac{4 \left(D t_1\right)^{n/2} \phi}{\Gamma\left(n/2 +1\right)}  \int^{\infty}_0 \frac{\left(1 - e^{- \rho}\right)}{\left(1 + e^{-  \rho}\right)^3} \, \exp\left(- \rho\right) \rho^n \, d\rho \, J_n(\rho) \,,
\end{align}
where $\phi = \sqrt{t_1/(t-t_1)} = \sqrt{z/(1-z)}$ (with $z =t_1/t$) and $J_n(\rho)$ is the following integral
\begin{align}
\label{J}
J_n(\rho) =  \frac{1}{\pi} \int^{1}_0 dx \left(1 - x^2\right)^{n/2} 
 \Bigg[\sinh(\rho) \frac{\cos\left(\phi x \rho\right)}{1+ \phi^2 x^2} +  \frac{\sin\left(\phi x  \rho\right)}{\phi x} +  
\cosh(\rho) \frac{\phi x  \, \sin\left( \phi x \rho\right)}{1+ \phi^2 x^2}\Bigg] \,.
\end{align}
 
It is expedient next to check the expression in eq. \eqref{kl} in the particular limit when $t_1 \to t$, i.e., when 
the auxiliary range ${\cal R}$ becomes the full range of BM on the interval $[0,t]$,  ${\cal R} \equiv R_t$. In this limit 
$\phi \to \infty$, the integral in eq. \eqref{J}  becomes
\begin{align}
J_n(\rho) \sim \frac{1}{2 \phi} \,,
\end{align}
in the leading in $\phi$ order, and hence
we recover from eq. \eqref{kl} the standard expression for the moments 
of the range of a Brownian motion up to time $t$:
 \begin{align}
\label{kl1}
 \mathbb{E}\left\{R_t^n\right\} =& \frac{4 \left(D t\right)^{n/2}}{\Gamma\left(1+n/2\right)} \int^{\infty}_0 \frac{\left(1 - e^{- \rho}\right)}{\left(1 + e^{-  \rho}\right)^3} \, \exp\left(- \rho\right) \rho^n \, d\rho = \nonumber\\
 =& \frac{4 \left(2^n - 4\right) n! \zeta\left(n - 1\right)}{2^n \Gamma\left(n/2 +1\right)} \left(D t\right)^{n/2} \,,
\end{align} 
 where $\zeta(n-1)$ is the Riemann zeta-function. Note that for $n=2$ this expression has to be understood as a limit $n \to 2$. In consequence, the integral over $\rho$ entering eq. (\ref{kl}) can be written down as
 \begin{equation}
 \int^{\infty}_0 \frac{\left(1 - e^{- \rho}\right)}{\left(1 + e^{-  \rho}\right)^3} \, \exp\left(- \rho\right) \rho^n \, d\rho = \frac{\Gamma\left(n/2+1\right)}{4 (D t)^{n/2}}  \mathbb{E}\left\{R_t^n\right\} = \frac{\left(2^n-4\right) n! \zeta(n-1)}{2^n} \,,
 \end{equation}
 which will permit us to express the moments of the auxiliary range in terms of the moments of the full range.
 
 To this end, we first notice that the expressions in the integrand in eq. (\ref{J}) admit the following expansion in the Taylor series in powers of $\rho$:
\begin{align}
&\sinh(\rho) \frac{\cos\left(\phi x \rho\right)}{1+ \phi^2 x^2} +    
\cosh(\rho) \frac{\phi x  \, \sin\left( \phi x \rho\right)}{1+ \phi^2 x^2} = \sum_{k=0}^{\infty} \frac{\rho^{2k+1}}{(2k+1)!}  \sum_{m=0}^k (-1)^m {2k \choose 2m} (x \phi)^{2 m}  \,,\nonumber\\
&\frac{\sin\left(\phi x  \rho\right)}{\phi x}  =  \sum_{k=0}^{\infty} \frac{(-1)^k \rho^{2k+1} (x \phi)^{2 k}}{(2k+1)!} \,.
\end{align}
Inserting these expansions into eq. (\ref{J}) and integrating over $x$, we arrive at the following expression for $J_n(\rho)$:
\begin{align}
\label{J1}
J_n(\rho) =& \frac{\Gamma(n/2+1)}{2 \pi} \Bigg[\sum_{k=0}^{\infty} \frac{(-1)^k \Gamma(k+1/2) \rho^{2k+1} \phi^{2 k}}{(2k+1)! \Gamma(k + n/2+3/2)} + \nonumber\\
+&
\sum_{k=0}^{\infty} \frac{\rho^{2k+1}}{(2k+1)!} 
\sum_{m=0}^k \frac{(-1)^m \Gamma(m+1/2)}{\Gamma(m+n/2+3/2)} {2k \choose 2m} \phi^{2m} \Bigg] \,.
\end{align}
Further on, plugging eq. (\ref{J1}) into eq. (\ref{kl7}), and integrating over $\rho$ we have
\begin{align}
\label{kl9}
& \mathbb{E}\left\{{\cal R}^n\right\}  = \frac{\left(t_1/t\right)^{n/2} \phi}{2 \pi \left(D t\right)^{3/2}} 
 \Bigg[ \sum_{k=0}^{\infty} \frac{(-1)^k \Gamma(k+1/2)}{(2k+1)!} \frac{\mathbb{E}\left\{R_t^{2 k + n +1}\right\}}{\left(D t\right)^{k}} \phi^{2 k} + \nonumber\\
+&
\sum_{k=0}^{\infty} \frac{\Gamma\left(k +(n+3)/2\right) \mathbb{E}\left\{R_t^{2 k + n +1}\right\}}{(2k+1)! (D t)^k} 
\sum_{m=0}^k \frac{(-1)^m \Gamma(m+1/2)}{\Gamma(m+n/2+3/2)} {2k \choose 2m} \phi^{2m} 
\Bigg] \,,
\end{align}
or, explicitly,
\begin{align}
\label{kl10}
&\mathbb{E}\left\{{\cal R}^n\right\}  = \frac{2 \left(D t_1\right)^{n/2} \phi}{\pi} \Bigg[\sum_{k=0}^{\infty} \frac{(-1)^k \left(2^{2 k+n}-2\right)(2 k + n+1)! \Gamma(k+1/2)}{2^{2 k+n} (2k+1)! \Gamma(k+(n+3)/2)} \zeta(2 k + n) \phi^{2 k} + \nonumber\\
 +&\sum_{k=0}^{\infty} \frac{\left(2^{2 k+n}-2\right)(2 k + n+1)!}{2^{2 k +n} (2 k+1)!} \zeta(2 k + n) 
\sum_{m=0}^k \frac{(-1)^m \Gamma(m+1/2)}{\Gamma(m+n/2+3/2)} {2k \choose 2m} \phi^{2m} 
\Bigg] \,.
\end{align}

We concentrate next on the moments of the zeroth and of the second order, i.e.,  
$n=0$ and $n=2$. For $n=0$ we have
\begin{align}
\label{kln0}
 \mathbb{E}\left\{{\cal R}^0\right\}  &= \frac{2 \phi}{\pi} \Bigg[2 + \sum_{k=1}^{\infty} \frac{(-1)^k \left(1-2^{1- 2 k}\right)}{(k + 1/2)} \zeta(2 k) \times \nonumber\\
 &\times \left( \phi^{2 k} + \frac{\left(\phi - i\right)}{2 \phi} \left(\phi - i\right)^{2 k} + \frac{\left(\phi + i\right)}{2 \phi} \left(\phi + i\right)^{2 k} \right)
\Bigg] \,. 
\end{align}
Using the integral representation of the zeta-function,
\begin{align}
\label{zeta}
\zeta(s) = \frac{1}{\left(1 - 2^{1-s}\right) \Gamma(s)} \int^{\infty}_0 \frac{x^{s-1} dx}{e^x +1} \,, 
\end{align}
which strictly holds for $Re \, s > 0$, we can perform both the summation in eq. \eqref{kln0} and 
also the integral over $dx$ to get
\begin{align}
\label{kln000}
\mathbb{E}\left\{{\cal R}^0\right\}  = \frac{2 \phi}{\pi} \Bigg[ 2+ g_0(\phi) + \frac{(\phi - i)}{2 \phi} g_0(\phi - i) + \frac{(\phi + i)}{2 \phi} g_0(\phi + i)
\Bigg] \,,
\end{align}
where
\begin{align}
\label{gg}
g_0(\phi) = \frac{\pi}{4 \phi} - 1 - 2 \, {\rm arccoth}\left(e^{\pi \phi}\right) + \frac{1}{2 \pi \phi} \left({\rm Li}_2\left(e^{- 2 \pi \phi}\right) - 4 \, {\rm Li}_2\left(e^{- \pi \phi}\right)\right) \,,
\end{align}
with ${\rm arccoth}(\ldots)$ being the inverse hyperbolic cotangent and
${\rm Li}_2(\ldots)$ - the Euler's dilogarithm. One can readily verify that  $\mathbb{E}\left\{{\cal R}^0\right\} \equiv 1$, as it should.

Next, for $n=2$ we have
\begin{align}
\label{kln2}
&\mathbb{E}\left\{{\cal R}^2\right\}  = \frac{2 D t_1 \phi}{\pi} \Bigg[4 \sum_{k=0}^{\infty} \frac{(-1)^k \left(1 - 2^{- 2 k - 1}\right) (k+1)}{(k+1/2)} \zeta(2 k + 2) \phi^{2 k} + \nonumber\\
+& \frac{1}{\phi^3} \sum_{k=0}^{\infty} \frac{(-1)^k \left(1 - 2^{- 2 k - 1}\right)}{(k+1/2)} \zeta(2 k + 2) \Bigg(
\left(2 (k+1) \phi - i\right) (\phi + i)^{2 k+2} + \nonumber\\
+&  \left(2 (k+1) \phi + i\right) (\phi - i)^{2 k+2}
\Bigg)
\Bigg] \,,
\end{align}
or, equivalently,
\begin{align}
\label{kln22}
\mathbb{E}\left\{{\cal R}^2\right\}  &= \frac{2 D t_1 \phi}{\pi} \Bigg(4 g_2(\phi) + \frac{2 (\phi+i)^2}{\phi^2} g_2(\phi+i) + 
\frac{2 (\phi-i)^2}{\phi^2} g_2(\phi-i)  + \nonumber\\
& + \frac{i}{\phi^3} \Big( (\phi - i)^2 g_1(\phi - i) - (\phi + i)^2 g_1(\phi+i)
\Big)\Bigg) \,,
\end{align}
where the function $g_1(\phi)$ is defined in eq. \eqref{gg1}, while
\begin{align}
\label{g1}
&g_2(\phi) = \frac{1}{2 \phi} \frac{d}{d\phi} \left(\phi^2 g_1(\phi)\right) = \frac{1}{2} g_1(\phi) + \frac{1}{2 \phi^2} \left(1 - \frac{\pi \phi}{\sinh(\pi \phi)}\right) \,.
\end{align}
Expressing $g_2(\phi)$ through $g_1(\phi)$ in eq. \eqref{kln22}, we find eventually
\begin{align}
\label{kln202}
\mathbb{E}\left\{{\cal R}^2\right\}  &= \frac{2 D t_1 \phi}{\pi} \Bigg(\frac{4}{\phi^2} + 2 g_1(\phi) + \frac{\left(1 + \phi^2\right)}{\phi^3}  \Big[(\phi + i) g_1(\phi+i)  + (\phi - i) g_1(\phi - i) 
\Big]
\Bigg) \,.
\end{align}
 
To access the small-$\phi$ behaviour of $\mathbb{E}\left\{{\cal R}^2\right\}$, it is necessary to work out a convergent small-$\phi$ representation of $g_1(\phi)$. This can be found from eq. \eqref{gg1} and reads 
\begin{align}
&g_1(\phi) = - \frac{2}{\phi} \sum_{k=1}^{\infty} \frac{(-1)^k}{k} {\rm arctg}\left(\frac{\phi}{k}\right) \,. \\
\end{align}
The latter representation in form of the expansion in the inverse tangents permits us to  
establish the following small-$\phi$ behaviour: 
\begin{align}
\label{small}
g_1(\phi) = \frac{\pi^2}{6} - \frac{7 \pi^4}{1080} \phi^2 + \frac{31 \pi^6}{75600} \phi^4 + O\left(\phi^6\right) \,,
\end{align}
such that
\begin{align}
(\phi - i) g_1(\phi - i) &+ (\phi + i) g_1(\phi+i)  = \pi + \left(\frac{4}{3} - \frac{\pi^2}{9}\right) \phi^3 + \nonumber\\
&+ \frac{\left(7 \pi^4 + 60 \pi^2 - 1440\right)}{900} \phi^5
 + O\left(\phi^7\right) \,.
\end{align}
In consequence, we find that for small values of the parameter $\phi$ (which corresponds to either small $t_1$ at a fixed $t$ or large $t$ at a fixed $t_1$)  $\mathbb{E}\left\{{\cal R}^2\right\} $ obeys
\begin{align}
\mathbb{E}\left\{{\cal R}^2\right\} &= 2 D (t - t_1)\Bigg[ 1 + \frac{4}{\pi} \phi + \phi^2 + \left(\frac{4}{3 \pi} + \frac{\pi}{18}\right) \phi^3 + O\left(\phi^4\right)\Bigg] \,.
\end{align}
Note that for $t_1 = 0$ the leading term in the latter equation, i.e., $2 D t$, is precisely the second moment of the running maximum of a Brownian trajectory \cite{1}.

In a similar way, we construct a small-$z$ expansion for $\mathbb{E}\left\{{\cal R}^2\right\} $. For $g_1(\phi)$ we get, for $z \to 0$,
\begin{align}
\label{azer1}
g_1(\phi) = \frac{\pi^2}{6} - \frac{7 \pi^4}{1080} z + \left(\frac{31 \pi^6}{75600} - \frac{7 \pi^4}{1080} \right) z^2 + O\left(z^3\right) 
\end{align}
and 
\begin{align}
\label{azer2}
&(\phi - i) g_1(\phi - i) + (\phi + i) g_1(\phi+i) = 2 \left(\pi - \arctan\left(\frac{2}{\phi}\right)\right) - \nonumber\\
&- 2 \sum_{k=2}^{\infty} \frac{(-1)^k}{k} \left[\arctan\left(\frac{\phi - i}{k}\right) + \arctan\left(\frac{\phi + i}{k}\right)\right] = \nonumber\\
& = \pi + \left(\frac{4}{3} - \frac{\pi^2}{9}\right) z^{3/2} + \left(\frac{21}{80} + \frac{495 - 360 \pi^2 + 28 \pi^4}{3600}\right) z^{5/2} + 
O\left(z^{7/2}\right)  \,.
\end{align}
Equations \eqref{azer1} and \eqref{azer2}, together with eq. \eqref{chief2}, yield the asymptotic expansion in eq. \eqref{chief3}.

To analyse the large-$\phi$ behaviour of $\mathbb{E}\left\{{\cal R}^2\right\}$,( which corresponds to the limit $t_1 \to t$), 
we have to proceed in a different way. To get an appropriate large-$\phi$ representation we formally
 represent the integral in eq. \eqref{g1} over the interval $[0,1]$ as the difference of the integral over the semi-infinite interval $[0,\infty)$ and the integral over $[1,\infty)$. The first integral can be performed explicitly and is equal to $\pi \ln(2)/\phi$, while the second one can be represented as an expansion in the incomplete upper gamma functions $\Gamma(0, (2 n +1) \pi \phi)$. This gives the following exact representation of the function $g_1(\phi)$:
\begin{align}
\label{as}
g_1(\phi) = \frac{\pi \ln(2)}{\phi} - \frac{1}{\phi^2} + \frac{2 \pi}{\phi} \sum_{n=0}^{\infty} \Gamma\left(0, (2 n+1) \pi \phi\right) \,.
\end{align}
Note that for large values of $\phi$ (or $n \to \infty$), the leading behaviour of the incomplete gamma function follows
\begin{align}
 \Gamma\left(0, (2 n+1) \pi \phi\right) \sim \frac{\exp\left(- (2 n + 1) \pi \phi\right)}{(2 n+1) \pi \phi} \,,
\end{align}
which means that the third term in the right-hand-side of eq. \eqref{as} is exponentially small and can be safely neglected. 
In this limit we obtain
\begin{align}
\label{as2}
\mathbb{E}\left\{{\cal R}^2\right\} &= 8 \ln(2) D t_1 \Bigg[ 1 + \frac{1}{2 \phi^2}  + O\left(e^{-\phi}\right)\Bigg] \,.
\end{align}
Note that for $t_1 = t$, i.e., for ${\cal R} = R_t$ and $\phi = \infty$, we recover from eq. \eqref{as2} the classical result $\mathbb{E}\left\{R_t^2\right\} = 8 \ln(2) D t$. 
The expansion in eq. \eqref{as2} gives directly the asymptotic form in eq. \eqref{chief4}.

\section{Conclusions}
\label{conc}

To conclude, we studied here the covariance function of the running range $R_t$ of a Brownian motion - one of extremal values of this paradigmatic stochastic process, which defines the maximal extent of a given Brownian trajectory on a time interval $[0,t]$. We
have calculated exactly this two-time autocorrelation function and analysed its asymptotic behaviour. Our analysis revealed a very non-trivial form of such a correlation function and also demonstrated that the temporal correlations 
between the values of the running range achieved on different time intervals
are very strong. 

As a by product of our analysis, we have determined the moments of an auxiliary range - the sum of the maximal displacement achieved on the entire time interval $[0,t]$ and of the partial minimum - the minimal displacement achieved on a smaller time interval $[0, t_1]$, $t_1 \leq t$. These complicated expressions and the corresponding probability density function of an auxiliary range will be discussed in our future work.

\ack

The authors acknowledge helpful discussions with Gr\'egory Schehr and Katja Lindenberg.

\end{document}